\documentclass[11pt,showpacs,onecolumn]{revtex4-1}
\usepackage{amsmath,amsfonts,amssymb}
\usepackage{graphicx}
\usepackage{algorithm}
\usepackage{algorithmic}

\newcommand{\arctanh}[1]{\mathrm{arctanh}#1}

\def\Tbethe{T_{\text{Bethe}}}
\def\Bbethe{\beta_{\text{Bethe}}}

\newcommand{\ud}{\mathrm{d}}

\newcommand{\Jij}{J_{i,j}}
\newcommand{\ie}{\textit{ i.e. }}
\newcommand{\eg}{\textit{ e.g. }}
\newcommand{\etal}{\textit{ et. al. }}
\newcommand{\Ham}{\mathcal{H}}
\newcommand{\B}{\beta}
\newcommand{\TCVM}{T_\text{CVM}}
\newcommand{\BCVM}{\B_\text{CVM}}
\newcommand{\BSI}{\B_\text{SG}}
\newcommand{\rref}[1]{eq. (\ref{#1})}

\newcommand{\<}{\langle}
\renewcommand{\>}{\rangle}
\newcommand{\cP}{\mathcal{P}}
\newcommand{\cL}{\mathcal{L}}
\newcommand{\cR}{\mathcal{R}}
\newcommand{\cU}{\mathcal{U}}
\newcommand{\cD}{\mathcal{D}}

\begin{document}

\title{Characterizing and Improving Generalized Belief Propagation
  Algorithms on the 2D Edwards-Anderson Model}

\author{Eduardo Dom\'\i nguez, Alejandro Lage-Castellanos, Roberto Mulet}
\affiliation{Department of Theoretical Physics and
  ``Henri-Poincar'e-Group'' of Complex Systems, Physics Faculty,
  University of Havana, La Habana, CP 10400, Cuba. }

\author{Federico Ricci-Tersenghi} \affiliation{Dipartimento di Fisica,
  INFN -- Sezione di Roma 1 and CNR -- IPCF, UOS di Roma,\\
  Universit\`{a} La Sapienza, P.le A. Moro 5, 00185 Roma, Italy}

\author{Tommaso Rizzo} \affiliation{Dipartimento di Fisica and CNR --
  IPCF, UOS di Roma,\\ Universit\`{a} La Sapienza, P.le A. Moro 5,
  00185 Roma, Italy}

\date{\today}

\begin{abstract}
We study the performance of different message passing algorithms in
the two dimensional Edwards Anderson model. We show that the standard
Belief Propagation (BP) algorithm converges only at high temperature
to a paramagnetic solution. Then, we test a Generalized Belief
Propagation (GBP) algorithm, derived from a Cluster Variational Method
(CVM) at the plaquette level. We compare its performance with BP and
with other algorithms derived under the same approximation: Double
Loop (DL) and a two-ways message passing algorithm (HAK). The
plaquette-CVM approximation improves BP in at least three ways: the
quality of the paramagnetic solution at high temperatures, a better
estimate (lower) for the critical temperature, and the fact that the
GBP message passing algorithm converges also to non paramagnetic
solutions. The lack of convergence of the standard GBP message passing
algorithm at low temperatures seems to be related to the
implementation details and not to the appearance of long range
order. In fact, we prove that a gauge invariance of the constrained
CVM free energy can be exploited to derive a new message passing
algorithm which converges at even lower temperatures. In all its
region of convergence this new algorithm is faster than HAK and DL by
some orders of magnitude.
\end{abstract}


\maketitle

\section{Introduction}

The 2D Edwards-Anderson (EA) model in statistical mechanics is defined
by a set $\sigma=\{s_1\ldots s_N\}$ of $N$ Ising spins $s_i=\pm 1$
placed on the nodes of a 2D square lattice, and random interactions
$\Jij$ at the edges, with a Hamiltonian
\[
\Ham (\sigma) = - \sum_{<i,j>} \Jij s_i s_j
\]
where $<i,j>$ runs over all couples of neighboring spins (first
neighbors on the lattice). The $\Jij$ are the magnetic interchange
constants between spins and are supposed fixed for any given instance
of the system, and the spins $s_i$ are the dynamic variables.  We will
focus on one of the most common disorder types, the bimodal
interactions $J=\pm 1$ with equal probabilities.
 
The statistical mechanics of the EA model, at a temperature $T=1/\B $,
is given by the Gibbs-Boltzmann distribution
\[P(\sigma) = \frac{e^{-\B\Ham(\sigma)}}{Z} \quad \mbox{where }\quad Z=\sum_{\sigma} e^{-\B \Ham(\sigma)}
\]
The direct computation of the partition function $Z$, or any marginal
probability distribution like $p(s_i,s_j)=\sum_{\sigma \backslash
  s_i,s_j}P(\sigma)$, is a time consuming task, unattainable in
general, and therefore an approximation is required. We are interested
in fast algorithms for inferring such marginal distributions. Actually
for the 2D EA model, thanks to the graph planarity, algorithms
computing $Z$ in a time polynomial in $N$ exist. However we are
interested in very fast (i.e.\ linear in $N$) algorithms that can be
used also for more general model, e.g.\ the EA model in a field or
defined on a 3D cubic lattice. For these more general cases a
polynomial algorithm is very unlikely to exist and some approximations
are required.

A simple and effective mean field approximation is the one due to
Bethe \cite{bethe}, in which the marginals over the dynamic variables,
like $p(s_i)$, are obtained from the minimization of a variational
free energy in a self consistent way. The Bethe approximation is exact
for a model without loops in the interactions network, which
unfortunately is far from being the usual case in physics. In the
context of finite dimensional lattices, Kikuchi \cite{kikuchi} derived
an extension of this approximation to larger groups of variables,
which accounts for short loops exactly, and is usually referred as
Cluster Variational Method (CVM).

The interest in spin glasses, with quenched random disorder, brought a
new testing ground for both approximations. In particular Bethe
approximation (exact on trees) has been the starting point of many
useful theoretical and applied developments. It is at the basis of the
cavity method, which allows a restatement of replica theory in
probabilistic terms for finite connectivity systems \cite{MP1}. The
Bethe approximation is connected to well known algorithms in computer
science, namely Belief Propagation \cite{pearlBP} and the sum-product
algorithm \cite{sumprod}. A major achievement of this confluence
between computer science and statistical mechanics, has been the
conception of the Survey Propagation algorithm \cite{KS,KS2}, inspired
by the cavity method and the replica symmetry breaking
\cite{MP1,MP2,MPV}, that shows great performance on hard optimization
problems \cite{KS,KS2,Col,Col2}. Statistical mechanics clarified the
relation between phase transitions and easy-hard transitions in
optimization problems, and allowed the statistical characterization of
the onset of the hard phase
\cite{achlioptas_rigorous,lenkaPNAS07,mon08}, as well as the
analytical description of search algorithms based on BP
\cite{MRTS_Allerton, BPdecimation}.

The correctness of Bethe approximation and the related algorithms is,
however, linked to the lack of topological correlations in the
interactions (random graphs are locally tree-like), since the
approximation is exact only on tree topologies. This is a strong
limitation for physical purposes, since tree topologies or random
graphs are not the common situation. Bethe approximation performs
poorly in finite dimensional lattices, and the associated algorithm
are usually non convergent at low temperatures.

Recently the Cluster Variational Method (CVM) has been reformulated in
a broader probabilistic framework called \textit{region-based}
approximations to free energy \cite{yedidia} and connected to a
Generalized Belief Propagation (GBP) algorithm to find the stationary
points of the free energy. It extends Bethe approximation by
considering correlations in larger regions, allowing, in principle, to
take into account short loops accurately. In \cite{yedidia} was shown
that stable fixed points of GBP message passing algorithm corresponds
to stationary points of the approximated CVM free energy, while the
converse is not necessarily true. Furthermore, the GBP message passing
is not guaranteed to converge at all. Prompted by this lack of
convergence, a new kind of provably convergent algorithms for
minimizing the CVM approximated free energy, known as Double Loop (DL)
algorithms \cite{yuille,HAK03}, has been developed, at the cost of a
drastic drop off in speed.

GBP has been applied in the last decade to inference problems
\cite{tanakaCVM,haploCVM,kappenCVMmedical}, consistently outperforming
BP. In particular, the image reconstruction problems
\cite{tanakaCVM1995,tanakaCVM} are based on a 2D lattices structure,
but, at variance with 2D EA model, the interactions among nearby spins
(pixels) are ferromagnetic, and the damaged image is used as an
external field. Both factors help convergence of GBP algorithms. An
analysis of CVM approximation using GBP algorithms on single instances
of finite dimensional disordered models of physical interest, like the
EA model, has not been done so far.

The Edwards Anderson model in 2D has been largely studied by other
methods (see \cite{JLMM, middleton09} and reference therein)
suggesting that it remains paramagnetic all the way down to zero
temperature, lacking any thermodynamic transition at any finite $T$,
although at low T there are metastable states of very long lifetime,
leading to very slow dynamics. Based on this fact, a paramagnetic
version of the GBP on 2D EA model was studied recently in
\cite{dual}. The connection of CVM with the replica trick and a
Generalized Survey Propagation have been presented recently
\cite{tommaso_CVM}. However the implementation of the latter algorithm
on finite dimensional lattices is computationally very demanding, and
should be preceded by the study of the original CVM approximation and
GBP algorithm.

In this paper we study the convergence properties of GBP message
passing algorithm and the performance of the CVM approximation on the
2D EA model. After the introduction of the region-based free energy in
Sec.~\ref{GBP2D} and the message passing algorithm in terms of cavity
fields, we compute the critical (inverse) temperature $\TCVM \simeq
0.82$ ($\BCVM \simeq 1.22$) of the plaquette-CVM approximation in
Sec.~\ref{Tc}, improving Bethe estimate $T_\text{Bethe}=1.51$
($\B_\text{Bethe} \simeq 0.66$) by roughly a factor 2. The CVM average
case temperature, however, does not clearly corresponds to the single
instance behavior of the GBP message passing algorithm, as is shown in
Sec.~\ref{ConvergenceProblems}. At variance with Belief Propagation,
GBP converges to spin glass solutions (below $T_\text{SG} \simeq
1.27$, above $\BSI \simeq 0.79$), and stops converging near $T \simeq
1.0$, before the average case prediction $\TCVM$. In Sec.\ref{gauge}
we show that this convergence problem depends on the implementation
details of the message passing algorithm, and can be improved by a
simultaneous update of message. In order to do so the gauge invariance
of the message passing equations has to be fixed. In
Sec.~\ref{GBPvsDL} we compare the solutions and the performance of GBP
with 3 other algorithms for the minimization of the CVM free energy:
Double Loop \cite{HAK03}, Two-Ways Message Passing \cite{HAK03}, and
the Dual algorithm \cite{dual}. In terms of the CVM free energy, the
paramagnetic solution is in general the one to be chosen, except for a
small interval in temperatures where the spin glass solution has a
lower free energy. Our results are summarized in
Sec.~\ref{Conclusions}.

\section{Generalized Belief Propagation on EA 2D}
\label{GBP2D}

Given that a detailed derivation of plaquette-GBP message passing
equations for the 2D Edwards Anderson model were presented in
\cite{dual}, here we only summarize such derivation, skipping
unnecessary details.

The idea of the \textit{region-based} free energy approximation
\cite{yedidia,pelizzola05} is to mimic the exact (Boltzmann-Gibbs)
distribution $P(\sigma)$, by a reduced set of its marginals. A
hierarchy of approximations is given by the size of such marginals,
starting with the set of all single spins marginals $p_i(s_i)$ (mean
field), then following to all neighboring sites marginals $p(s_i,s_j)$
(Bethe approximation), then to all square plaquettes marginals
$p(s_i,s_j,s_k,s_l)$, and so on. Since the only way of knowing such
marginals exactly is the unattainable computation of $Z$, the method
pretends to approximate them by a set of beliefs $b_i(s_i)$,
$b_L(s_i,s_j)$, $b_\cP(s_i,s_j,s_k,s_l)$, etc. obtained from a
minimization of a region based free energy.

Following the derivation done in \cite{dual}, the plaquette level
approximated free energy for the 2D EA model is given as a
contribution of all Plaquettes, Links and Spins in the 2D lattice:
\begin{widetext}
\begin{eqnarray}
-\B F &=& \sum_{\cP} \sum_{\sigma_\cP} \displaystyle  b_\cP(\sigma_\cP) \log \frac{ b_\cP(\sigma_\cP)}{\exp(-\B E_\cP(\sigma_\cP))} \qquad\mbox{Plaquettes} \nonumber\\
& & -\sum_{L} \sum_{\sigma_L} \displaystyle  b_L(\sigma_L) \log \frac{ b_L(\sigma_L)}{\exp(-\B E_L(\sigma_L))}   \qquad\mbox{Links}  \label{eq:freeen} \\
& & +\sum_{i} \sum_{s_i} \displaystyle  b_i(s_i) \log \frac{ b_i(s_i)}{\exp(-\B E_i(s_i))}  \qquad\mbox{Spins} \nonumber
 \end{eqnarray}
\end{widetext}
where the symbol $\sigma_R=(s_1,\ldots,s_k)$ stands for the set of
spins in region $R$, while $E_R(\sigma_R)=-\sum_{<i,j>\in R} \Jij s_i
s_j$ stands for the energy contribution in that region. The energy
term $E_i(s_i)$ in the spins contribution is only relevant when an
external field acts over spins, and will be neglected from now on.

\begin{figure}[htb]
\begin{center}              
\includegraphics[width=0.6\textwidth]{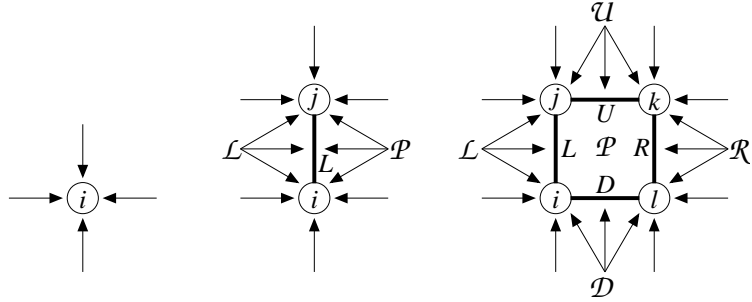}
\caption{Schematic representation of belief equations
  (\ref{eq:beliefs}).  Lagrange multipliers are depicted as arrows,
  going from parent regions to children regions. Italics capital
  letters are used to denote Plaquettes, simple capital letters denote
  Links, and lower case letters denote Spins.}
\label{fig:beliefs2D}
\end{center} 
\end{figure}

An unrestricted minimization of the free energy (\ref{eq:freeen}) in
terms of its beliefs, produces incongruent results. Beliefs are only
meaningful as an approximation to the correct marginals if they obey
the marginalization constrains $b_i(s_i) = \sum_{s_j} b_L(s_i,s_j)$
and $b_L(s_i,s_j) = \sum_{s_k,s_l} b_P(s_i,s_j,s_k,s_l)$. This
marginalization is enforced by the introduction of Lagrange
multipliers (see \cite{yedidia} for a general introduction, and
\cite{dual} for this particular case) in the free energy
expression. There is one Lagrange multiplier $\mu_{L\to i}(s_i)$ for
every link $L$ and spin $i \in L$, and a Lagrange multiplier
$\nu_{\cP\to L}(s_i,s_j)$ for each plaquette $\cP$ and link $L\in\cP$
. In terms of these Lagrange multipliers, the stationary condition of
the approximated free energy is achieved with
\begin{eqnarray}
b_i(s_i) &=& \frac{1}{Z_i} \exp\left(-\B E_i(s_i) - \sum_{L\supset
  i}^4 \mu_{L\to i}(s_i)\right)\;, \nonumber\\
b_L(\sigma_L) &=& \frac{1}{Z_L} \exp\left(-\B E_L(\sigma_L) -
\sum_{\cP \supset L}^2 \nu_{\cP\to L}(\sigma_L) - \sum_{i\subset L}^2
\mathop{\sum_{L'\supset i}^3}_{L' \neq L} \mu_{L'\to i}(s_i)
\right)\;, \label{eq:beliefs}\\
b_\cP(\sigma_\cP) &=& \frac{1}{Z_\cP} \exp\left(-\B E_\cP(\sigma_\cP)
- \sum_{L\subset\cP}^4 \mathop{\sum_{\cP'\supset L}^1}_{\cP' \neq \cP}
\nu_{\cP'\to L}(\sigma_L) - \sum_{i\subset \cP}^4 \mathop{\sum_{L \supset
    i}^2}_{L \not\subset \cP} \mu_{L\to i}(s_i) \right)\;. \nonumber
\end{eqnarray}
A graphical representation of these equations is given in figure
\ref{fig:beliefs2D}. Lagrange multipliers are shown as arrows going
from parent regions, to children. Take, for one, the middle equation
for the belief in link regions $b_{L}(\sigma_L)=b_{L}(s_i,s_j)$. The
sum of the two Lagrange multipliers $\nu_{\cP\to L} (s_i,s_j)$
corresponds to the triple arrows on both sides of the link in central
figure \ref{fig:beliefs2D}, while the two sums over three messages
$\mu_{L'\to i}(s_i)$ corresponds to the three arrows acting over the
top ($j$) and bottom ($i$) spins, respectively.  In equations
(\ref{eq:beliefs}), the $Z_R$ are normalization constants. The terms
$E_\cP(\sigma_\cP)=E_\cP(s_i,s_j,s_k,s_l)=-(J_{i,j} s_i s_j+J_{j,k}
s_j s_k+J_{k,l} s_k s_l+J_{l,i} s_l s_i)$ and $E_L(s_i,s_j)=- J_{i,j}
s_i s_j$ are the corresponding energies in plaquettes and links
respectively, and are represented in the diagram by the lines
(interactions) between circles (spins). 
zero since no field is acting upon spins.

\begin{figure}[htb]
\begin{center}              
\includegraphics[width=0.6\textwidth]{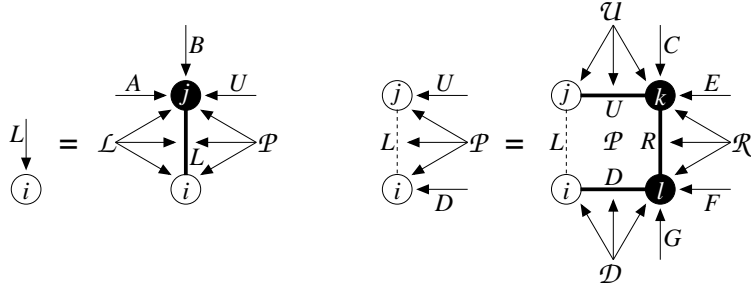} 
\caption{Message passing equations (\ref{eq:message-u}) and
  (\ref{eq:message-Uuu}), shown schematically.  Messages are depicted
  as arrows, going from parent regions to children regions. On any
  link $\Jij$, represented as bold lines between spins (circles), a
  Boltzmann factor $e^{\B \Jij s_i s_j}$ exists. Dark circles
  represent spins to be traced over. Messages from plaquettes to links
  $\nu_{P\to L}(s_i,s_j)$ are represented by a triple arrow, because
  they can be written in terms of three parameters $U$, $u_i$ and
  $u_j$, defining the correlation $\< s_i s_j \>$ and magnetizations
  $\< s_i \>$ and $\< s_j \>$, respectively.}
\label{fig:message2D}
\end{center} 
\end{figure}

The Lagrange multipliers can be parametrized in terms of cavity
fields $u$ and $(U,u_a,u_b)$ as
\begin{eqnarray}
-\mu_{L\to i}(s_i) &=& \B u_{L\to i} \: s_i \\
-\nu_{\cP \to L}(s_i,s_j)&=& \B (U_{\cP \to L} \:s_i s_j + u_{\cP \to i} \:s_i + u_{\cP \to j}\:  s_j)
\end{eqnarray}
In particular, the field $u_{L\to i}$ corresponds to the cavity field
in the Bethe approximation \cite{yedidia}. The choice of these
parametrization is the reason for the use of single and triple arrows
in figures \ref{fig:beliefs2D} and \ref{fig:message2D}. In particular,
the messages going from plaquettes to links, are characterized by
three fields ($U_{\cP \to L},u_{\cP \to i},u_{\cP \to j}$), and the
capital $U_{\cP \to L}$ acts as an effective interaction term.

The Lagrange multipliers are related among them by the constrains they
are supposed to impose (see \cite{dual}). In terms of the cavity
fields and using the notation in figure \ref{fig:message2D},
Link-to-Spin cavity fields shall be related by
\begin{equation}
u_{L\to i} = \hat{u}(u_{\cP\to i} + u_{\cL \to i},\;
U_{\cP\to L} + U_{\cL\to L} + J_{ij},\;
u_{\cP\to j} + u_{\cL\to j} + u_{A\to j} + u_{B\to j} + u_{U\to j})\;,
\label{eq:message-u}
\end{equation}
where
\[
\hat{u}(u,U,h) \equiv u + \frac{1}{2\B} \log\frac{\cosh\B(U+h)}{\cosh\B(U-h)}
\]
Note that the usual cavity equation for fields in the Bethe
approximation \cite{MP1} is recovered if all contributions from
plaquettes $\cP$ and $\cL$ are set to zero.

Similarly, by imposing the marginalization of the beliefs at
Plaquettes onto their children Links, we find the self consistent
expression for the Plaquette-to-Link cavity fields:
\begin{eqnarray}
U_{\cP\to L} &=& \hat U(\#) =  \frac{1}{4\B}\log\frac{K(1,1)K(-1,-1)}{K(1,-1)K(-1,1)} \nonumber \\
u_{\cP\to i} &=& - u_{D\to i}  + \hat u_i(\#) =u_{\cD\to i} - u_{D\to i} + \frac{1}{4\B}\log\frac{K(1,1)K(1,-1)}{K(-1,1)K(-1,-1)}
\label{eq:message-Uuu} \\
u_{\cP\to j} &=& - u_{U\to j} + \hat u_j(\#) = u_{\cU\to j} - u_{U\to j} + \frac{1}{4\B}\log\frac{K(1,1)K(-1,1)}{K(1,-1)K(-1,-1)} \nonumber
\end{eqnarray}
where
\begin{eqnarray*}
K(s_i,s_j) &=& \sum_{s_k,s_l} \exp \bigg[
\B \Big( (U_{\cU\to U} + J_{jk}) s_j s_k +
(U_{\cR\to R} + J_{kl}) s_k s_l + (U_{\cD\to D} + J_{li}) s_l s_i + \\
&& (u_{\cU\to k} + u_{C\to k} + u_{E\to k} + u_{\cR\to k}) s_k +
(u_{\cR\to l} + u_{F\to l} + u_{G\to l} + u_{\cD\to l}) s_l \Big) \bigg]
\end{eqnarray*}
and the symbol $\#$ stands for all incoming fields in the right hand
side of the equations. The functions $\hat u(u,U,h)$ and $[\hat
  U(\#),\hat u_i(\#),\hat u_j(\#)]$ will be used in next section for
the average case calculation.

For a given system of size $N$ (number of spins) there are $2N$ Links
and $N$ square plaquettes, and therefore there are $4 N$
Plaquette-to-Link fields $[U_{\cP\to L},u_{\cP\to i},u_{\cP\to j}]$,
and $4 N$ Link-to-Spins fields $u_{L\to i}$. At the stationary points
of the free energy their values are related by the set of $4N+4N$
equations (\ref{eq:message-u}) and (\ref{eq:message-Uuu}).

The set of $4 N +4 N$ self-consistent equations are also called
message-passing equations when they are used as update rules for
fields in the message passing algorithm, or cavity iteration equations
in the context of cavity calculations. The field notation is more
comprehensible than the original Lagrange multipliers notation, and
has a clear physical meaning: each plaquette is telling its children
links that they should add an effective interaction term $U_{P\to L}$
to the direct interaction $\Jij$, due to the fact that spins $s_i$ and
$s_j$ are also interacting through the other three links in the
plaquette. Terms $u_i$ act like magnetic fields upon spins, and the
complete $\nu(s_i,s_j)-$message is characterized by the triplet
$(U_{i,j},u_i,u_j)$.

\section{Critical Temperature of Plaquette-CVM approximation}
\label{Tc}
In this section we revisit the method used in \cite{tommaso_CVM} to
compute the critical temperature at which CVM approximation develops a
spin glass phase. By spin glass phase we mean a phase characterized by
non zero local magnetizations $m_i = \tanh \left(\B \sum_L^4 u_{L\to
    i} \right)$ and nearly zero total magnetization $m = \frac 1 N
\sum_i m_i \simeq 0$ (remember we work with no external field). The 2D
EA model is paramagnetic down to zero temperature, but spin glass like
solutions can appear in the CVM approximation due to its mean field
character.  We correct one of the conclusions reached in
\cite{tommaso_CVM}, where we fail to observe the appearance of the
spin glass phase in the CVM approximation to the 2D Edwards Anderson
model. We follow an average case approach, which is similar in spirit
but different from the single instance stability analysis done in
\cite{MooKap05} for the Bethe approximation (Belief Propagation).

The average case calculation is a mathematical technique developed in
\cite{MP1} to study the typical solutions of cavity equations in
disordered systems, with a deep and fundamental connection to the
replica trick \cite{MPV}. When applied to the plaquette-CVM
approximation \cite{tommaso_CVM}, we end up with two equations, in
which fields (messages) are now replaced by functions of fields $q(u)$
and $Q(U,u_1,u_2)$, and the interactions are averaged out. As a
consequence of the homogeneity of the 2D lattice and the averaging
over local disorder $\Jij$, all plaquettes, links, and spins in the
graph are now equivalent, and we only need to study one of them to
characterize the whole system.

More precisely, the average case self consistent equations for the
distribution $q(u)$ is given by
\begin{eqnarray}
q(u_i) &=& \mathbb{E}_J \int \ud q(u_{A\to j})\: \ud q(u_{B\to j}) \: \ud q(u_{U\to j}) \label{eq:ave_qu_eq}    \\
&& \ud Q(U_{\cP \to L},u_{\cP \to i},u_{\cP \to j})  \: \ud Q(U_{\cL \to L},u_{\cL \to i},u_{\cL \to j})\; \delta\Bigl(u_i - \hat{u}(\#)\Bigr)  \nonumber
\end{eqnarray}
with $\hat u(\#)$ as defined in the right hand side of equation (\ref{eq:message-u}), and $\ud f(x) \equiv f(x) \ud x$

The corresponding self-consistent equation for $Q(U,u_1,u_2)$ is
\begin{eqnarray}
\iint  Q(U,u_a,u_b) q(u_i-u_a) q(u_j-u_b) \ud u_a \ud u_b = \phantom{\iint  Q(U,u_1,u_2)} \label{eq:ave_QUuu_eq} \\
 = \mathbb{E}_J \int \ud q(u_{C\to k})  \: \ud q(u_{E\to k}) \: \ud q(u_{F\to l}) \: \ud q(u_{G\to l}) \: \ud Q(U_{\cU\to U},u_{\cU \to j},u_{\cU \to k})\: \nonumber \\ 
\ud Q(U_{\cR\to R},u_{\cR \to k},u_{\cR \to l}) \: \ud Q(U_{\cD\to D},u_{\cD \to l},u_{\cD \to i}) 
\delta\Bigl(U-\hat U (\#)\Bigr) \; \delta\Bigl(u_i-\hat u_i (\#)\Bigr) \; \delta\Bigl(u_j-\hat u_j (\#)\Bigr) \nonumber 
\end{eqnarray}
where the notation corresponds to equation (\ref{eq:message-Uuu}). In
both equations (\ref{eq:ave_qu_eq}) and (\ref{eq:ave_QUuu_eq}) the
expression $\mathbb{E}_J = \int \ud J P(J) \ldots$ stands for the
average over the quenched randomness.

At high temperatures we expect fixed point equations
(\ref{eq:message-u}) and (\ref{eq:message-Uuu}) to yield a
paramagnetic solution. Such a solution is characterized by Link to
Site messages $u=0$, and Plaquette to Link messages $(U,u_1,u_2) =
(U,0,0)$. If we impose this ansatz to fields, we recover the
paramagnetic or dual algorithm of \cite{dual} for the single instance
message passing, and the paramagnetic average case study of
\cite{tommaso_CVM} for the average case. Let us remember that the 2D
EA model is expected to have no thermodynamic transition at any finite
temperature, and hence remain paramagnetic all the way down to
$T=0$. Following \cite{tommaso_CVM}, in the average case the
paramagnetic solution has the form
\begin{eqnarray*}
 q(u) &=& \delta(u) \\
Q(U,u_1,u_2) &=& Q(U) \delta(u_1)\delta(u_2)
\end{eqnarray*}
The equation (\ref{eq:ave_qu_eq}) is always satisfied when
$q(u)=\delta(u)$ for whatever $Q(U)$. The equation
(\ref{eq:ave_QUuu_eq}) can be solved self-consistently for $Q(U)$:
\begin{eqnarray} Q(U) &=& \mathbb{E}_J \int \ud Q(U_{\cU})\: \ud Q(U_{\cR})\:\ud Q(U_{\cD})  \label{eq:ave_QU} \\
&& \delta\left(U - \frac{1}{\B} \arctanh\Bigl[\tanh \B(J_\cU + U_\cU)\tanh \B (J_\cR + U_\cR)\tanh \B (J_\cD + U_\cD)\Bigr]\right) \nonumber
\end{eqnarray}
and the average free energy and all other relevant functions can be
derived in terms of it (see \cite{tommaso_CVM}).

On the other hand, a general (not paramagnetic) solution of the
average case equations (\ref{eq:ave_qu_eq}) and (\ref{eq:ave_QUuu_eq})
is very difficult, since it involves the deconvolution of
distributions $q(u)$ in the left hand side of \rref{eq:ave_QUuu_eq} in
order to update $Q(U,u_1,u_2)$ by an iterative method. A critical
temperature can be found, however, by an expansion in small $u$ around
the paramagnetic solution. We can focus on the second moments of the
distributions
\begin{eqnarray*}
 a &=& \int q(u) u^2 \ud u \\
 a_{i\:j}(U) &=& \iint Q(U,u_1,u_2) \:u_i\: u_j \:\ud u_1 \ud u_2  \qquad \text{where } i,j \in \{1,2\} 
\end{eqnarray*}
and check whether the paramagnetic solution ($a=0$ and $a_{ij}(U) =
0$) is locally stable. To do this we expand equations
(\ref{eq:ave_qu_eq}) and (\ref{eq:ave_QUuu_eq}) to second order, and
we obtain the following linearized equations:
\begin{eqnarray*}
 a &=& K_{a,a} a + \int \ud U'\: K_{a,a_{11}}(U') a_{11}(U') + \int \ud U'\: K_{a,a_{12}}(U') a_{12}(U') \\ 
a\: Q(U) +a_{11}(U) &=& K_{a_{11},a}(U) a + \int \ud U'\: K_{a_{11},a_{11}}(U,U') a_{11}(U') + \int \ud U'\: K_{a_{11},a_{12}}(U,U') a_{12}(U') \\ 
a_{12}(U) &=& K_{a_{12},a}(U) a + \int \ud U'\: K_{a_{12},a_{11}}(U,U') a_{11}(U') + \int \ud U'\: K_{a_{12},a_{12}}(U,U') a_{12}(U')  
\end{eqnarray*}
The actual values of the $K_{a_x,a_y}$ come from the expansion in
small $u$ of the original equations (see equation 90 in
\cite{tommaso_CVM} for an example).

We can not solve these equations analytically because we do not have
an analytical expression of $Q(U)$ for the paramagnetic solution at
all temperatures. By discretizing the values of $U$ uniformly in
$(-U_{\text{max}},U_{\text{max}})$, \ie $U= i \Delta U$ with
$i\in[-I_{\text{max}},I_{\text{max}}]$, we can transform the
continuous set of equations to a system of the form
\begin{equation}
\vec a = \mathbf{K}(\B) \cdot \vec a \label{eq:matrix_a_Ka} 
\end{equation}
where the vector of the second moments $\vec a=
(a,a_{11}(U),a_{12}(U))$ have the form
\begin{eqnarray*}
\vec a &=& \bigl(a, a_{11}(-U_\text{max}),a_{11}(-U_\text{max}+\Delta U ),\ldots,a_{11}(U_\text{max}-\Delta U ),a_{11}(U_\text{max}), \bigr.  \\
&& \bigl.\phantom{\bigl(a, }\: a_{12}(-U_\text{max}),a_{12}(-U_\text{max}+\Delta U ),\ldots,a_{12}(U_\text{max}-\Delta U ),a_{12}(U_\text{max}) \bigr) 
\end{eqnarray*}
$\mathbf{K}(\B)$ is a $(2 I_\text{max}+1) \times (2 I_\text{max}+1)$
matrix, that stand for the discrete representation of the integrals in
the right hand side of the linearized equations, and depends on the
inverse temperature via the solution $Q(U)$ of eq. (\ref{eq:ave_QU}).

The paramagnetic solution $\vec a =0$ always satisfy the homogeneous
\rref{eq:matrix_a_Ka}. The stability criterion for the paramagnetic
solution is the singularity of the Jacobian $\det (\mathbf
I-\mathbf{K}(\B)) =0$. When such condition is satisfied, a non
paramagnetic solution continuously arises from the paramagnetic one,
since a flat direction appears in the free energy.

\begin{figure}[!htb]
\includegraphics[angle=0,width=0.5\textwidth]{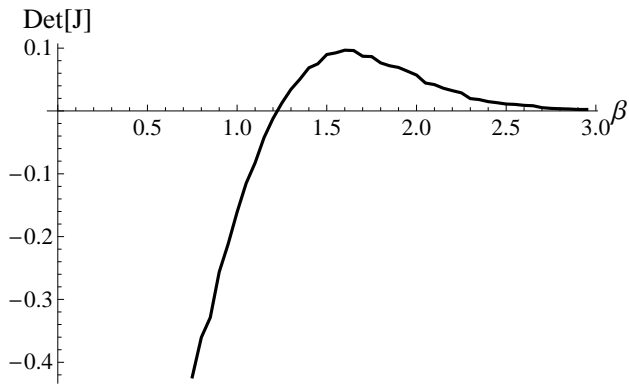}
\caption[0]{Determinant of the Jacobian $\mathbf J = \mathbf
  I-\mathbf{K}(\B)$ as a function of inverse temperature $\B$. The
  critical inverse temperature is $\BCVM \simeq 1.22$.}
\label{fig:eigen_B_c}
\end{figure}

Numerically, we worked with a discretization of $2 I_\text{max} + 1 =
41$ points between $(- U_\text{max} = -3.5,\:U_\text{max} = 3.5
)$. The paramagnetic solution $Q(U)$ is found solving
eq. (\ref{eq:ave_QU}) by an iterative method at every temperature, and
then used to compute the elements of the $\mathbf{K}(\B)$ matrix. In
figure \ref{fig:eigen_B_c} we show the determinant of the Jacobian
matrix $\mathbf J = \mathbf I - \mathbf K(\B)$. The critical inverse
temperature derived from this analysis is $\BCVM \simeq 1.22$ for the
appearance of a flat direction in the free energy.

In \cite{tommaso_CVM} $\BCVM$ was thought to be infinite (zero
temperature) because an incomplete range of the values of $\B$ was
examined. The critical temperature found here is below the Bethe
critical temperature $\B_{\text{Bethe}} \simeq 0.66$, and therefore
improves the Bethe approximation by roughly a factor 2, since the 2D
EA model is likely to remain paramagnetic at all finite
temperatures. At variance with the Bethe approximation, the single
instance behavior of the message passing is not so clearly related to
the average case critical temperature, as we show in the next section.

\section{Performance of GBP on 2D EA Model}
\label{ConvergenceProblems}

\begin{figure}[tb]
\includegraphics[angle=270,width=0.8\textwidth]{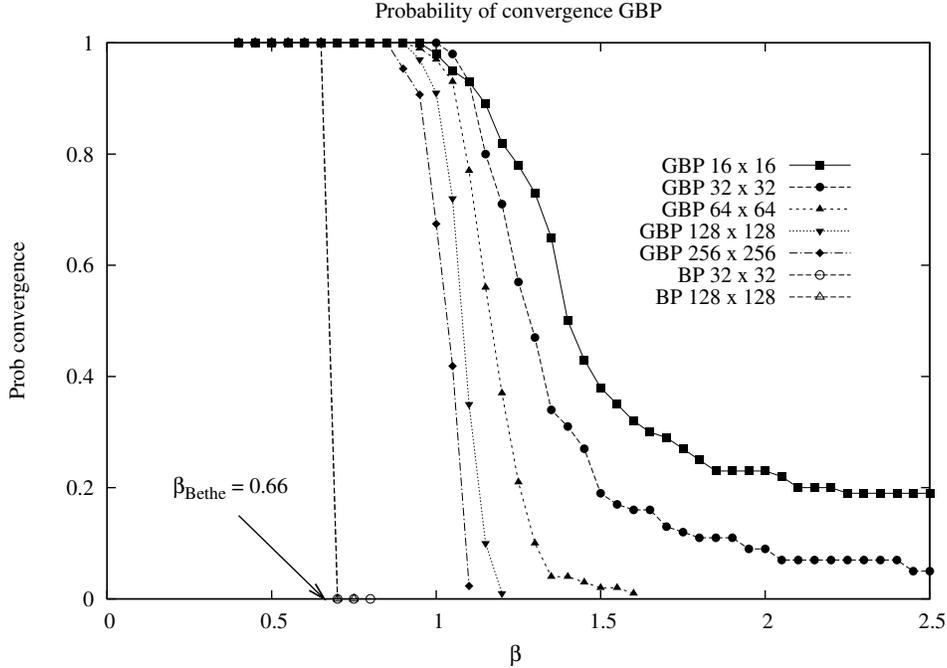}
\caption[0]{Probability of convergence of BP and GBP on a 2D EA model,
  with random bimodal interactions, as a function of inverse
  temperature $\B = 1/T$. The Bethe spin glass transition is expected
  to occur at $\Bbethe\simeq 0.66$ on a random graph with the same
  connectivity. The BP message passing algorithm on 2D EA model stops
  converging very close to that point. Above that temperature, BP
  equations converge to the paramagnetic solution, \ie all messages
  are trivial $u=0$. Below the Bethe temperature (nearly) the Bethe
  instability takes messages away from the paramagnetic solution, and
  the presence of short loops is thought to be responsible of the lack
  of convergence. On the other hand, the GBP equations converge at
  lower temperatures, but eventually stops converging as well.}
\label{fig:prob_conv_GBP-BP}
\end{figure}

Before studying GBP message passing for the plaquette-CVM
approximation, let us check what happens to the simpler Bethe
approximation and the corresponding message passing algorithm known as
Belief Propagation (BP) in the 2D EA model. When running BP at high
temperatures (above $\Tbethe = 1/\B_{\text{Bethe}} \simeq 1.51$) in a
typical instance of the model with bimodal interactions, we find the
paramagnetic solution (given by all fields $u=0$), and therefore, the
system is equivalent to a set of independent interacting pairs of
spins, which is only correct at infinite temperature. The Bethe
temperature $\Tbethe$ (computed in average case and exact on acyclic
graphs \cite{defTbethe}), seems to mark precisely the point where BP
stops converging (see Fig. \ref{fig:prob_conv_GBP-BP}). Indeed
messages flow away from zero below $\Tbethe$, and convergence of the
BP message passing algorithm is not achieved anymore. So, the Bethe
approximation is disappointing when applied to single instances of the
Edwards Anderson model: either it converge to a paramagnetic solution
at high temperatures, or it does not converge at all below
$T_\text{Bethe}$.

The natural question arises, as to what extent GBP message passing
algorithm for the plaquette-CVM approximation is also non convergent
below its critical temperature, and whether this temperature coincides
with the average case one. To check this we used GBP message passing
equations (\ref{eq:message-u}) and (\ref{eq:message-Uuu}), with a
damping factor $0.5$ in the Link-to-Site fields $u$:
\[u^{\text{new}}_{L\to i} = 0.5 \: u^{\text{old}}_{L\to i} + 0.5\:
\hat u (\#) \] We will make the distinction between two types of
solutions for the GBP algorithm. The high temperature or paramagnetic
solution is characterized by zero local magnetization of spins $m_i =
\sum_{s_i} s_i b_i (s_i) = \tanh \left(\B \sum_L^4 u_{L\to i}\right) =
0$. At low temperatures, following the average case analysis, a non
paramagnetic or spin-glass solution should appear, characterized by
non zero local magnetizations, but roughly null global
magnetization. The temperature at which non zero local magnetizations
appear will be called $T_\text{SG} = 1/\BSI$.

\begin{figure}[tb]
\includegraphics[angle=270,width=0.8\textwidth]{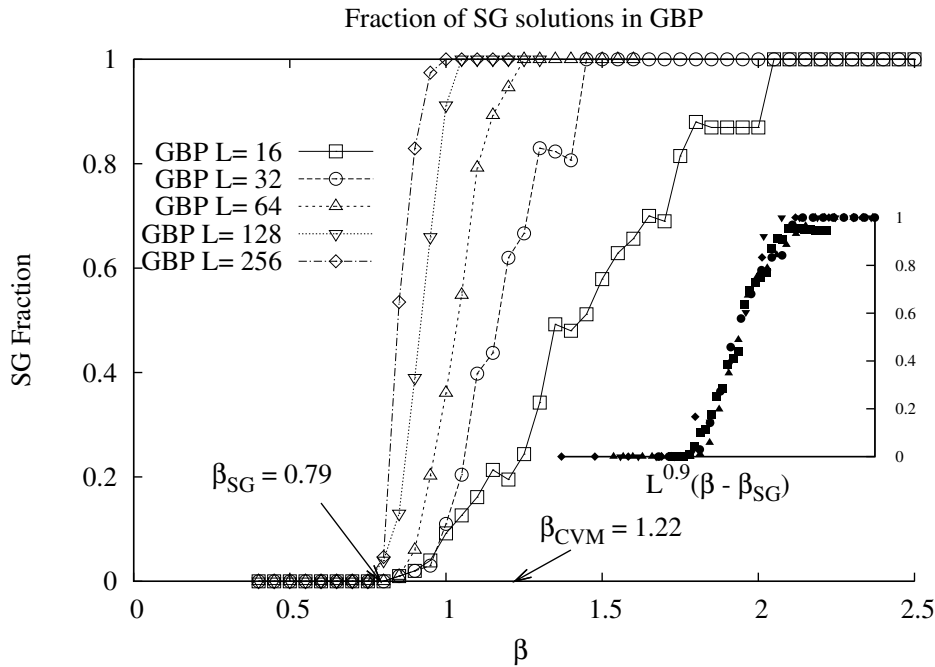}
\caption[0]{Data points correspond to the fraction of SG solutions in
  a population of 100 systems of sizes $16^2$, $32^2$, $64^2$,
  $128^2$, $256^2$ respectively. At high temperatures (low $\B$) GBP
  message passing converge always to the paramagnetic solution. The
  average case critical inverse temperature $\BCVM\simeq 1.22 $ does
  not corresponds to the single instance behavior, as the spin glass
  solutions in GBP appear around $\BSI \simeq 0.79$. The inset shows
  that all data collapsed if plotted as a function of the scaling
  variable $L^{0.9} (\B-0.79)$, where the exponent $0.9$ and the
  critical inverse temperature $\BSI \simeq 0.79$ are obtained from
  best data collapse.}
\label{fig:sg_frac}
\end{figure}

Figure \ref{fig:prob_conv_GBP-BP} shows that GBP is able to converge
below the Bethe critical temperature, but stops converging before the
CVM average case critical temperature $\BCVM\simeq 1.22$. Furthermore,
figure \ref{fig:sg_frac} shows that even before stop converging, GBP
finds a spin glass solution in most instances.

The inner plot of figure \ref{fig:sg_frac} shows a collapse of the
data points for different system sizes using the scaling variable
$L^{0.9} (\B-0.79)$, which gives an estimate $\BSI\simeq 0.79$ (the
exponent $0.9$ is obtained from the best data collapse).  Since
$\BSI\simeq 0.79$ is well below the average case inverse critical
temperature $\BCVM \simeq 1.22$, the relevance of the latter on the
behavior of GBP on single samples is questionable.  By a similar data
collapse procedure, we estimate the non-convergence temperature for
the GBP message passing algorithm to be $\B_{\text{conv}} \simeq 0.96$
(see Fig.~\ref{fig:GBPMat_B_c}), which is again far away from the
average case prediction $\BSI$.

So, beyond the simple Bethe approximation, we found three different
temperatures in the CVM approximation: $\BSI\simeq 0.79 <
\B_{\text{conv}} \simeq 0.96 < \BCVM \simeq 1.22$ corresponding
respectively to the appearance of spin glass solutions, to the lack of
convergence on single instances, and to the average case prediction
for the critical temperature.

We can summarize three main differences between the properties of BP
and GBP. At high temperatures (below $\BSI \simeq 0.79$) GBP gives a
quite good approximate of the marginals \cite{dual}, namely the
paramagnetic solution with non trivial correlations fields $U\neq 0$,
while BP treats the system as a set of independent pairs of linked
spins. Furthermore, this naive approach is all that BP can do for us,
since above $\B_\text{Bethe}\simeq 0.66$, it no longer converges. GBP,
on the other hand, is not only able to converge beyond
$\B_\text{Bethe}$, but it is also able to find spin glass solutions
above $\BSI$. The third difference between both algorithms is that the
non convergence of BP seems to occur exactly at the same temperature
where a spin glass phase should appear (and arguably because of it),
while the GBP convergence problems appear deep into the spin glass
phase. The lack of convergence of GBP, however, seems to depend
strongly on implementation details as we show next.

\section{Gauge invariance of GBP equations}
\label{gauge}

The convergence properties of the GBP message passing is sensitive to
implementation details, \eg the damping value in the update equations,
and this is not an inherent property of the CVM (or region-graph)
approximation. We might try, for instance, to update simultaneously
all \textit{small-}u fields pointing towards a given spin, hoping to
gain some more stability in message passing algorithm. When trying to
do this we find out that there is a freedom in the choice of these
fields that has no effect over the fixed point solutions. This freedom
(similar to the one noticed in \cite{martin_protein}) is the result of
having introduced unnecessary Lagrange multipliers to enforce
marginalization constraints that were already indirectly enforced.

\begin{figure}[!htb]
\includegraphics[angle=0,width=0.5\textwidth]{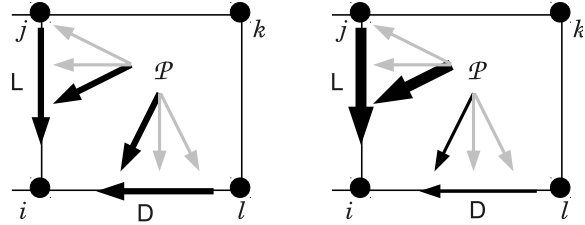}
\caption[0]{Null modes of the plaquette CVM free energy in terms of
  fields. The small-$u$ fields that act over a given spin $i$ inside a
  plaquette can be shifted by an arbitrary amount $\delta$ as in
  equation (\ref{eq:gauge_transfor}) without changing the self
  consistent (message passing) equations.}
\label{fig:invariance}
\end{figure}

Consider, for instance, the messages shown in figure
\ref{fig:invariance}. If the belief on a plaquette
$b_P(s_i,s_j,s_k,s_l)$ correctly marginalizes to the beliefs of two of
its children links $b_{L}(s_i,s_j)$ and $b_{D}(s_l,s_i)$, and one of
those beliefs marginalizes to the common spin $b_i(s_i)= \sum_{s_j}
b_{L}(s_i,s_j)$, it is inevitable that the second link $D$ also
marginalizes to the same belief on $s_i$, since $b_i(s_i) = \sum_{s_j}
b_{L}(s_i,s_j) = \sum_{s_j,s_l,s_k} b_{\cP}(s_i,s_j,s_k,s_l) =
\sum_{s_l} b_{D}(s_l,s_i)$. Therefore the Lagrange multiplier that was
introduced to force this last marginalization is not needed. This
redundancy is a general feature of GBP equations when there are more
than two level of regions (Plaquette, Links, and Spins, in our case).

The consequence of having introduced unnecessary multipliers leads to
a gauge invariance on the fields (messages) values. Such invariance
can be better understood by looking at the GBP equations at infinite
temperature: for $\B =0$ the non linear parts of the message passing
equations (\ref{eq:message-u}) and (\ref{eq:message-Uuu}) disappear,
but there is still a set of linear equations to be satisfied for the
small-$u$ messages with infinite many non trivial solutions. These
solutions correspond, however, to the same physical paramagnetic
solution, since the total field $h_i=\sum_L^4 u_{L\to i}$ and the
magnetizations $m_i = \tanh(\B h_i) $ are always zero. It is easy to
check that once we have a solution of the message passing equations
(\ref{eq:message-u}) and (\ref{eq:message-Uuu}) at any temperature, we
can change by an arbitrary amount $\delta$ any group of 4 $u$-messages
inside a plaquette (figure \ref{fig:invariance}) pointing to the same
spin as
\begin{eqnarray}
u_{L\to  i}&\to& u_{L\to i}+\delta\:, \nonumber\\
u_{\cP_L\to i}&\to& u_{\cP_L\to i}+\delta\:, \label{eq:gauge_transfor}\\
u_{D\to  i}&\to& u_{D\to i}-\delta\:,  \nonumber \\
u_{\cP_D\to i}&\to& u_{\cP_D\to i}-\delta\:,  \nonumber 
\end{eqnarray}
and still all self-consistent equations are satisfied.  

\begin{figure}[!htb]
\includegraphics[width=0.8\textwidth]{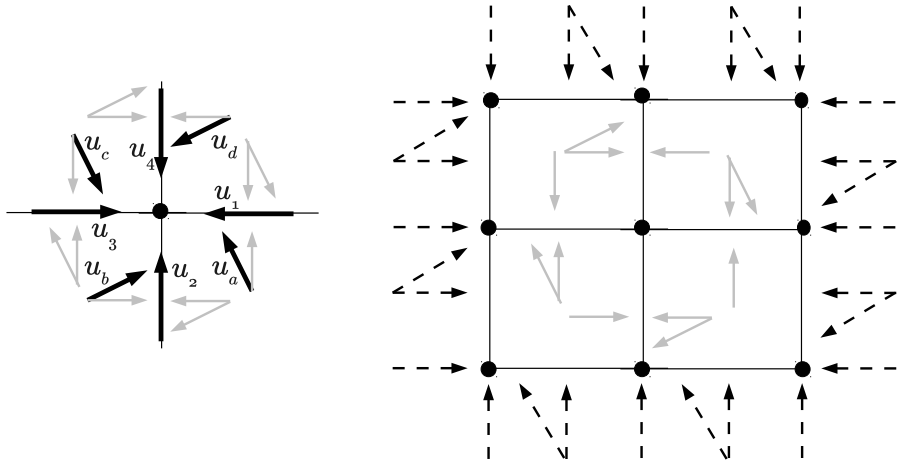}
\caption[0]{In the left diagram, all 8 \textit{small-}$u$ messages
  pointing to the central spin are highlighted with bold face. They
  are 4 Link-Site $u$-messages, and 4 Plaquette-Link
  $u_{\text{left}}$-messages. They have linear dependence among
  them. The right diagram shows four plaquettes around a spin, and the
  messages that contribute in a non linear way to the aforementioned 8
  messages. The idea of GBP+GF is to compute the non linear
  contributions to the message passing equations, and then assign the
  values of the $u$-messages in order to satisfy their linear
  relations.}
\label{fig:gauge_fixed_matrix}
\end{figure}

This local null mode of the standard GBP equations can be avoided by
arbitrary setting to zero one of the four small-$u$ fields entering
equation (\ref{eq:gauge_transfor}). We choose to fix the gauge by
removing the right small-$u$ field in every Plaquette-to-Link field
$(U,u_{\text{left}},u_{\text{right}})$, as shown in figure
\ref{fig:gauge_fixed_matrix}. Once the gauge is fixed, the fields are
uniquely determined, and we can try to implement the simultaneous
updating of all \textit{small-}u fields around a given spin, hopefully
improving convergence.

In the left diagram of figure \ref{fig:gauge_fixed_matrix} all
messages involving the central spin are represented, and in bold face
those that act precisely upon that spin. These messages enter linearly
in the message passing equations of each other (see equations
(\ref{eq:message-u}) and (\ref{eq:message-Uuu})). Therefore, the self
consistent equations they should satisfy at the fixed points, can be
written as (using the notation of figure \ref{fig:gauge_fixed_matrix})
\begin{equation}
\begin{array}{ll}
\begin{array}{ll}
 u_1 &= u_a + NL_1 \\
 u_2 &= u_b + NL_2 \\
 u_3 &= u_c + NL_3 \\
 u_4 &= u_d + NL_4 
\end{array} & \quad
\begin{array}{ll}
 u_a &=  u_b - u_2 + NL_a \\
 u_b &=  u_c - u_3 + NL_b \\
 u_c &=  u_d - u_4 + NL_c \\
 u_d &=  u_a - u_1 + NL_d 
\end{array}
\end{array} \label{eq:linear_system}
\end{equation}
where the $NL$ stand for the non linear contributions to the
corresponding equation. As a consequence, the values of the 8
$u$-messages pointing to the central spin can be assigned precisely by
a linear transformation for any given values of the non linear
contributions.  This gauge fixed updating method, that we will call
GBP+GF, updates all $u$-messages around a spin simultaneously and in a
way that they are consistent with each other via the message passing
equations.

\begin{figure}[!htb]
\includegraphics[angle=270,width=0.8\textwidth]{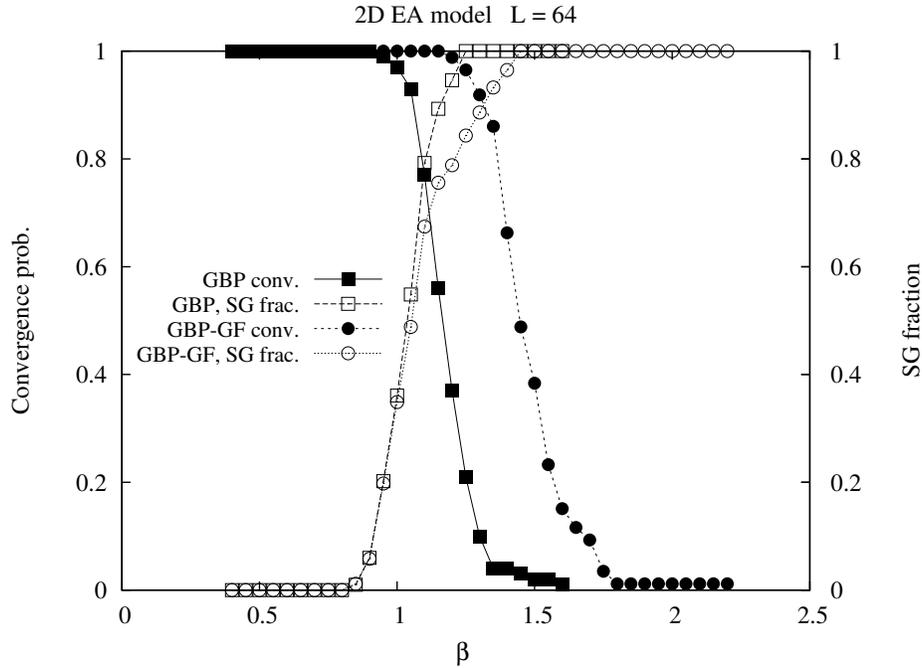}
\caption[0]{Convergence probability of GBP and GBP+GF as a function of
  $\B$. The solution found by either iteration method is always the
  same (when both converge), but GBP+GF reaches lower temperatures
  while converging. The fraction of spin glass solutions found by
  either algorithm show that GBP+GF sees the same spin glass
  transition temperature. The fraction of spin glass solutions is
  always given respect to the amount of convergent solutions.}
\label{fig:prob_conv_MatGBP_64}
\end{figure}

The right diagram in figure \ref{fig:gauge_fixed_matrix} shows the
messages entering the non linear parts. Taking the 8 $u$-messages as
zero, the non linear contributions are the right hand sides of the
message passing equations involved. With the non linear parts
computed, the system of equations (\ref{eq:linear_system}) is solved
for the $u$-variables multiplying the non linearities vector by the
corresponding matrix. The 8 $u$-messages are then updated, usually
with a damping factor. The update of the $U$ correlation fields is
done as in the original GBP method, via the equation
(\ref{eq:message-Uuu}), since it does not depend on the $u$-messages
that are being updated.

\begin{figure}[!htb]
\includegraphics[angle=270,width=0.7\textwidth]{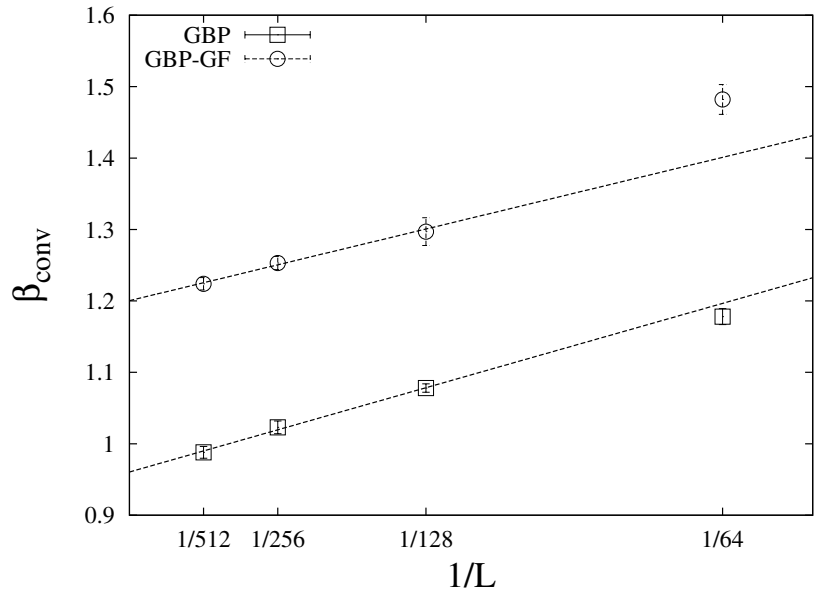}
\caption[0]{Estimate of the non convergence temperature for different
  system sizes using the standard GBP (squares) and the Gauge Fixed
  GBP (circles). As shown, with the gauge fixed procedure the non
  convergence extrapolated temperature is quite close to the average
  case prediction $\BCVM \simeq 1.22$. Each data point corresponds to
  the average of the non convergence temperature over many
  realizations of the disorder: 10 realizations for the $512 \times
  512$ systems, 20 for the $256 \times 256$ and 100 for the others.}
\label{fig:GBPMat_B_c}
\end{figure}

Figure \ref{fig:prob_conv_MatGBP_64} shows the probability of
convergence versus inverse temperature for GBP and GBP+GF, and also
the fraction of the solutions found that correspond to a spin glass
phase. Let us emphasize here that GBP and GBP+GF are not different
approximations, but different methods to find the same fixed point
solution by message passing. They are expected to find the same
solutions, and in fact they do. At high temperatures both methods
converge to the paramagnetic solution, with all null local
magnetizations $m_i=\tanh \left(\B\sum_L^4 u_{L\to i} \right)=0$. The
standard message passing update of GBP equations hardly converges
above $\B_\text{conv} \simeq 0.96$, while the GBP+GF method reaches
lower temperatures, $\B_{\text{conv-GF}}\simeq 1.2$, as can be seen in
Fig.~\ref{fig:GBPMat_B_c}. Furthermore, the GBP+GF allows us to work
in a range of temperatures where most solutions are spin glass
like. This proves that the non converging temperature found for GBP,
$\B_{\text{conv}}\simeq 0.96$, is not a feature of the CVM
approximation, but a characteristic of the message passing method
used, and can be outperformed by other message passing schemes, like
GBP+GF. Kindly note in figure \ref{fig:GBPMat_B_c} that the non
convergence inverse temperature of GBP+GF $\B_{\text{conv-GF}}\simeq
1.2$ is quite close to the average case prediction for the critical
temperature $\BCVM \simeq 1.22$. Whether this is accidental or not is
still unclear. Since the average case instability should describe the
breakdown of the paramagnetic phase, and the lack of convergence in
single instances occurs while already in a non paramagnetic phase, it
seems far fetched assuming that both critical behaviors are related.

\subsection{Gauge fixed average case stability}

The disagreement between the average case critical temperature $\BCVM$
and the one observed in the single instance $\BSI$, can be due to a
number of reasons. First, the average case calculation assumes that
cavity fields are uncorrelated. But, in our case, messages
participating in the cavity iteration are very close to each other in
the lattice, and thus correlated. Furthermore, GBP does not have the
equivalent of a Bethe lattice for BP, i.e.\ a model in which the
correlation between cavity messages is close to zero by
construction. The second reason for a failure of the average case
prediction is that the transition we observe in single instances might
be due to the almost inevitable appearance of ferromagnetic domains in
large systems (Griffith instability). The third, and the most obvious
reason, is that the gauge invariance was not accounted in the average
case calculation.

Reproducing the method of Sec.~\ref{Tc} to obtain an average case
prediction of the critical temperature for the Gauge Fixed GBP is not
straightforward. The reason is that Link-to-Spins messages $u$, should
fulfill two different equations: their own original equation
(\ref{eq:message-u}), and the implicit equation derived from the fact
that the gauge is fixed and one of the fields in the Plaquette-to-Link
message $(U,u,u)$ is set to zero.

\begin{figure}[!htb]
\includegraphics[angle=0,width=0.6\textwidth]{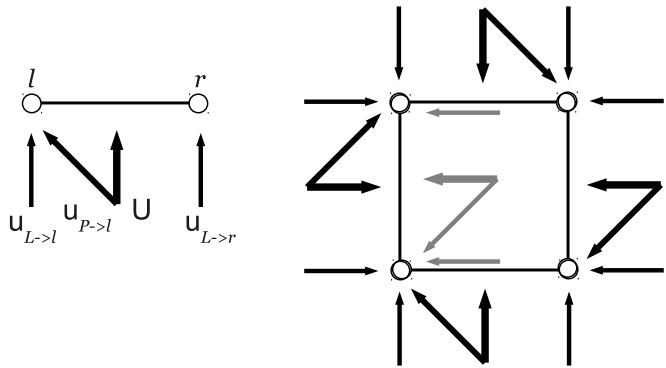}
\caption[0]{Left: The set of four messages that we compute jointly by
  a population dynamic. Right: the population dynamic step consists in
  taking four quadruplets at random from the population (those in
  black), and computing a new quadruplet (the one in gray inside the
  plaquette) using randomly selected interactions $J_{ij}$ on the
  plaquette.}
\label{fig:ave_case_pop_diagram}
\end{figure}

However, a different average case calculation is possible. We can
represent the messages flowing in the lattice by a population of
quadruplets $(u_{L_l\to l},u_{\mathcal P \to l},U_{\mathcal{P}\to lr},
u_{L_r\to r})$, where one of the original messages is absent because
the gauge has been fixed (see left panel in
Fig.~\ref{fig:ave_case_pop_diagram}). Given any four of these
quadruplets of messages around a plaquette, we can compute, using the
message passing equations, the new messages inside the plaquette (see
right panel in Fig.~\ref{fig:ave_case_pop_diagram}). The new
population dynamics consists in picking four of these quadruplets out
of the population at random, then computing the new quadruplet (using
also random interactions in the plaquette) and finally put it back in
the population. After several steps, the population stabilizes either
to a paramagnetic solution (where all $u=0$ and only $U\neq 0$),
either to a non paramagnetic one (where also $u \neq 0$).

\begin{figure}[tb]
\includegraphics[angle=270,width=0.7\textwidth]{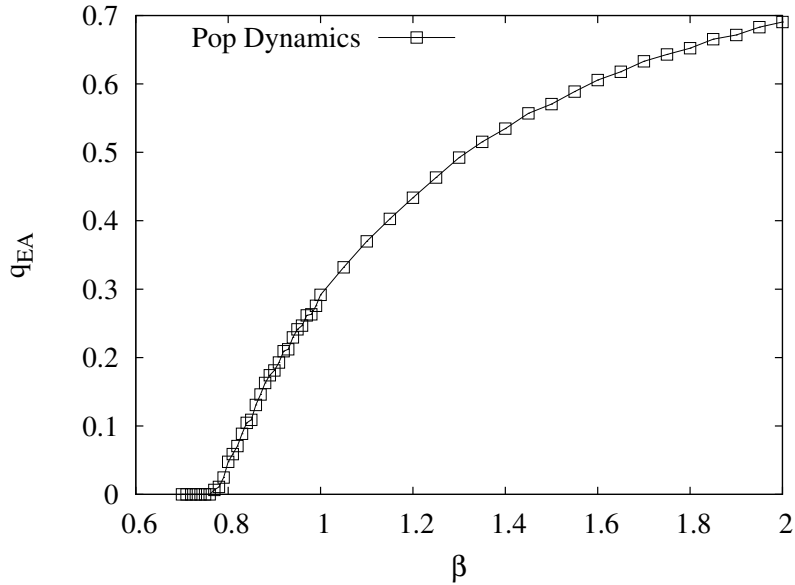}
\caption[0]{Edwards Anderson order parameter, see eq.(\ref{eq:qEA}),
  obtained using a population of $N=10^3$ messages, and running the
  population dynamic step $10^3 \times N$ times. In agreement with the
  single instance behavior, the transition between paramagnetic
  ($q_{EA} =0$) and non paramagnetic (spin glass) phases is found at
  $\B \simeq 0.78$.}
\label{fig:ave_case_pop}
\end{figure}

In Fig. \ref{fig:ave_case_pop} we show the Edwards Anderson order
parameter $q_{EA} = \sum_i m_i^2 / N$ obtained at different
temperatures using this population dynamics average case method. We
find that $q_{EA}$ becomes larger than zero at $\beta_\text{CVM-GF}
\simeq 0.78$, which is quite close to the inverse temperature $\BSI
\simeq 0.79$ where single instances develop non-zero local
magnetizations and a spin glass phase. The correspondence between this
average case result and the single instance behaviour is very
enlightening: indeed the average case computation does not take into
account correlations among quadruplets of messages and it is not
sensible to Griffith's singularities. So, the most simple explanation
for the GBP-GF behaviour on single samples of the 2D EA model is that
quadruplets of messages arriving on any given plaquette are mostly
uncorrelated and that at $\BSI$ a true spin glass instability takes
place (which is an artifact of the mean-field like approximation).
Please consider that under the Bethe approximation the SG instability
happens at $\beta_\text{Bethe} \simeq 0.66$, while the CVM
approximation improves the estimate of the SG critical boundary to
$\BSI \simeq 0.79$ (on single instances) and to $\beta_\text{CVM-GF}
\simeq 0.78$ (on the average case).

\section{Same approximation, four algorithms}
\label{GBPvsDL}

It can be proved \cite{yedidia} that stable fixed points of the
message passing equations correspond to stationary points of the
region graph approximated free energy (or CVM free energy). The
converse is not necessarily true, and some of the stationary points of
the free energy, might not be stable under the message passing
heuristic. As we have seen, the message passing might not even
converge at all. For a given free energy approximation
(eq. (\ref{eq:freeen}) in our case), there are other algorithms to
search for stationary points, including other types of message passing
and provably convergent algorithms. In this section we study two of
these algorithms and show that they do find the same spin glass like
transition at $\B_m$, but have a different behavior at lower
temperatures.

The one presented so far is the so called Parent-to-Child (PTC)
message passing algorithm, in which Lagrange multipliers are
introduced to force marginalization of bigger (parent) regions onto
their children. Other choices of Lagrange multipliers are possible
\cite{yedidia}, leading to the so called Child-to-Parent and Two-Ways
algorithms.  Next we test the following four algorithms for minimizing
the plaquette-CVM free energy in typical instances of 2D EA:
\begin{itemize}
 \item Double-Loop algorithm of Heskes \etal \cite{HAK03}. Is a
   provably convergent algorithm that guarantees a step by step
   minimization of the free energy functional. It consist of two
   loops, the inner of which is a Two-Ways message passing algorithm
   that we will call HAK. We use the implementation in LibDai public
   library \cite{libdai0.2.3}.
\item HAK message passing algorithm. Is a Two-Ways message passing
  algorithm \cite{HAK03}. When it converges, it is usually faster than
  Double-Loop.
\item GBP Parent-to-Child is the message passing algorithm we have
  presented so far in this paper, and for which the simultaneous
  updating of cavity fields was introduced to help
  convergence. Nevertheless the following results were obtained using
  standard GBP PTC.
\item Dual algorithm of \cite{dual}. Is the same GBP PTC setting all
  small fields $u=0$, and doing only message passing in terms of
  correlation fields $U$ (first equation in
  eq. (\ref{eq:message-Uuu})).
\end{itemize}

For the last three algorithms we use our own implementation in terms
of cavity fields $u$ and $(U,u_a,u_b)$. The dual algorithm forces the
solution of GBP to remain paramagnetic since all $u=0$. This
paramagnetic ansatz is specially suited for the 2D EA model since it
is expected to be paramagnetic at any finite temperature (in the
thermodynamical limit).

\begin{figure}[tb]
\includegraphics[angle=270,width=0.8\textwidth]{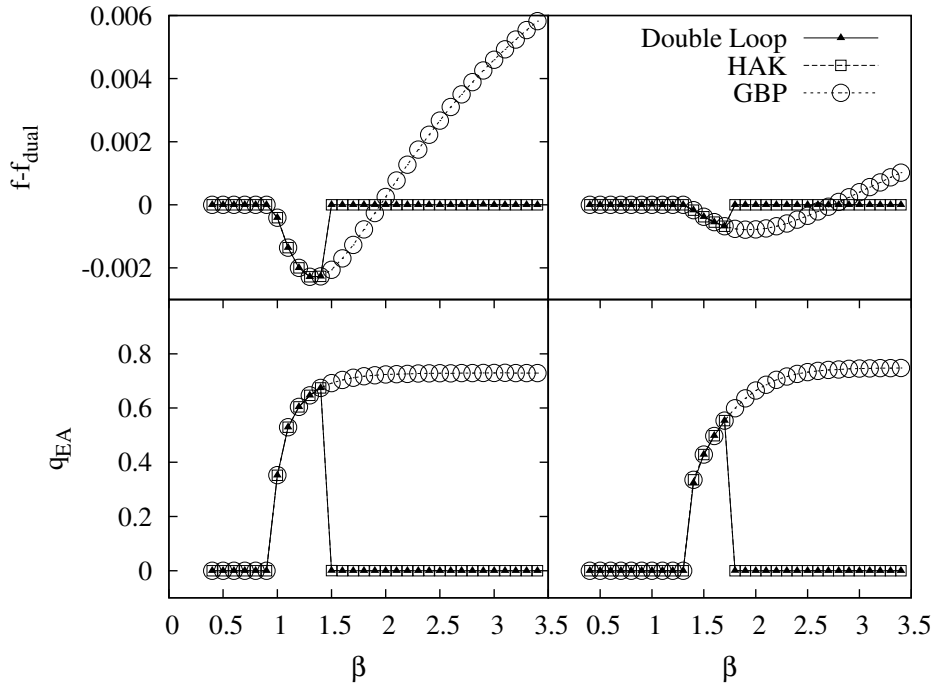}
\caption[0]{Free energy of the solutions found by Double Loop
  algorithm, HAK and the GBP PTC algorithm relative to the free energy
  of the paramagnetic solution (Dual approximation), in a typical
  system in which GBP PTC finds a spin glass solution. At high
  temperatures both algorithm find the same paramagnetic
  solution. Interestingly, there is a small range of temperatures
  where the spin glass solution found by GBP is actually the one that
  minimizes the free energy. But at even lower temperatures the
  paramagnetic solution becomes again the correct one. While Double
  Loop and HAK switch back to the paramagnetic solution (even if at a
  wrong $T$), the GBP PTC get stuck in the spin glass solution (and
  for this reason, it eventually stops converging).}
\label{fig:free}
\end{figure}

As shown in the previous section, the GBP PTC message passing
equations finds a paramagnetic solution in the 2D EA model at high
temperatures, while below $T_\text{SG}=1/\BSI \simeq 1.27$ it finds a
spin glass like solution. By spin glass like we mean that the total
field $h_i=\sum_L^4 u_{L\to i}$ and the magnetization $m_i = \tanh(\B
h_i)$ are non zero and change from spin to spin. The order parameter
\begin{equation}
q_{\text{EA}} = \frac{1}{N} \sum_{i} m_i^2 \label{eq:qEA}
\end{equation}
is used to locate this phase. The critical temperature $T_\text{SG}$,
where $q_\text{EA}$ becomes larger than zero, seems to be independent
of message passing details, like damping or the use of gauge fixing
for simultaneous updates of fields.

In figure \ref{fig:free} we show the free energy and the
$q_{\text{EA}}$ parameter of the solutions found by Double Loop, HAK
and GBP PTC for two typical realizations of an $N=16\times 16$ EA
system with bimodal interactions. The free energy of the dual
approximation is subtracted to highlight the differences with respect
to the paramagnetic solution. The figure shows that HAK and Double
Loop do find the same spin glass solution that GBP PTC finds when
going down in temperature. This solution is actually lower in free
energy when it appears, but at even lower temperatures becomes
subdominant compared to the paramagnetic one. The GBP PTC keeps
finding the spin glass solutions while Double Loop and HAK switch back
to the paramagnetic one. This is an interesting feature of Double Loop
and in particular of HAK which is a fast message passing algorithm. By
returning to the dual (paramagnetic) solution, HAK is also ensuring
its convergence at low temperature \cite{dual}, while GBP PTC get lost
in the irrelevant (and physically wrong) spin glass solution, and
eventually stops converging.

However note that DL and HAK may stop finding the SG solution when
this solution is still the one with lower free energy.  Moreover the
lack of convergence of GBP can be used as a warning that something
wrong is happening with the CVM approximation, something that is
impossible to understand by looking at the behavior of provably
convergent algorithms.

\begin{figure}[tb]
\includegraphics[angle=270,width=0.8\textwidth]{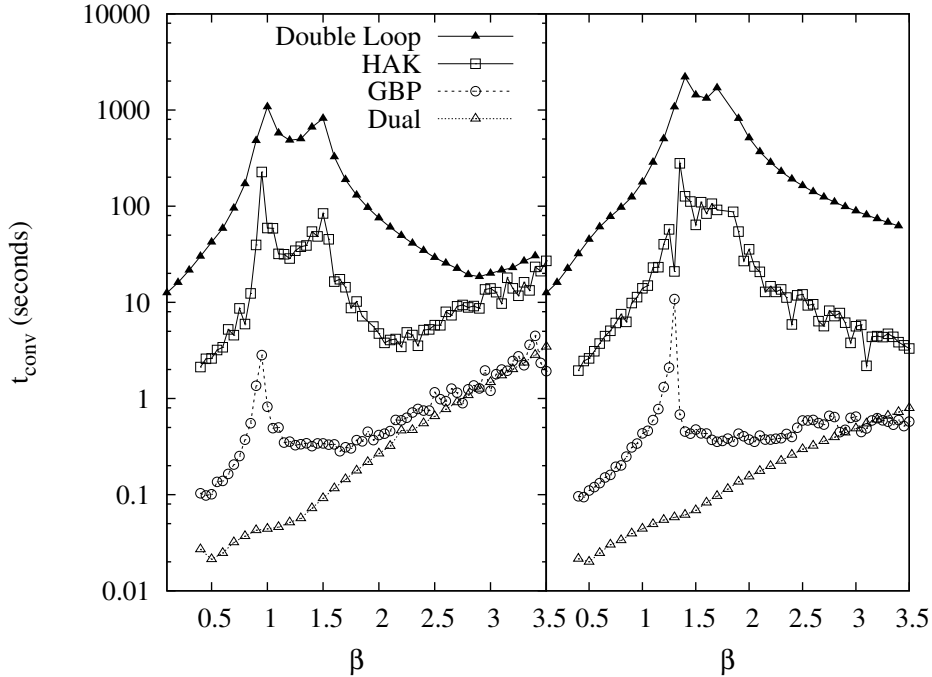}
\caption[0]{Convergence time in seconds for the Double Loop algorithm
  (full points) and standard message passing algorithms (empty points)
  for the plaquette-GBP approximation in two different realizations of
  a $16^2$ Edwards Anderson system. Message passing algorithms are
  typically faster, but not always convergent. The first cusp is
  related to the appearance of the spin glass solution, while the
  second cusp in the Double Loop algorithm is related to the switching
  back to the paramagnetic solution (see Fig. \ref{fig:free}).}
\label{fig:t_conv_libdai_GBP}
\end{figure}

In figure \ref{fig:t_conv_libdai_GBP} we compare the running times of
Double Loop (LibDai \cite{libdai0.2.3}), HAK and GBP PTC (our
implementation) for the two systems of figure \ref{fig:free}. As
expected, Double Loop is much more slowly than the message passing
heuristics of HAK and GBP (please notice the log scale in the time
axis). The peaks in running times correspond to the transition points
from paramagnetic to spin glass solution. Double Loop and HAK have two
peaks, the second corresponding to the transition back to paramagnetic
solution, while the GBP PTC has only the first peak.

\section{Summary and Conclusions}
\label{Conclusions}

We studied the properties of the Generalized Belief Propagation
algorithm derived from a Cluster Variational Method approximation to
the free energy of the Edwards Anderson model in 2D at the level of
plaquettes. We compared the results obtained by Parent-to-Child GBP
with the ones obtained by the Dual (paramagnetic) algorithm
\cite{dual} and by HAK Two-Ways algorithm \cite{HAK03} and Double-Loop
provably convergent algorithm \cite{HAK03}.

We found that the plaquette-CVM approximation (using Parent-to-Child
GBP) is far richer than the Bethe (BP) approximation in 2D EA
model. BP converges only at high temperatures (above
$T_{\text{Bethe}}=1/\B_{\text{Bethe}} =1.51$), and in such case it
treats the system as a set of independent pairs of linked spins.  GBP
on the other hand, makes a better prediction on the paramagnetic
behavior of the model at high T, since it implements a message passing
of correlations fields flowing from plaquettes to links in the
graph. Furthermore with GBP the paramagnetic phase is extended to
temperatures below $T_\text{Bethe}=1.51$ until $T_\text{SG} = 1/\BSI
\simeq 1.27$ where spin glass solutions appear in the single instance
implementation of the message passing algorithm. In contrast to Bethe
approximation, GBP is able to find spin glass solutions, and the
standard message passing stops converging near $T_{\text{conv}} \simeq
1$.

The average case calculation of the stability of the paramagnetic
solution in the CVM approximation predicted that non paramagnetic
(spin glass) solutions should appear at lower temperatures $ \TCVM =
1/\BCVM \simeq 0.82$. This average case result does not coincide with
the single instance behavior of the standard GBP, since it fails to
mark both the point where GBP start finding spin glass solutions
$T_\text{SG}$ and the point where GBP stops converging
$T_\text{conv}$.

However, the non convergence of GBP is not a feature of the CVM
approximation, and is susceptible of changes from one implementation
of the message passing to another. We showed that by fixing a hidden
gauge invariance in the message passing equation, a simultaneous
update of all cavity fields pointing to a single spin in the lattice
improves the convergence of the algorithm, without changing
drastically its speed. Using the gauge fixed GBP, the non
convergence inverse temperature is moved to $T_{\text{conv-GF}} \simeq
1.2$, quite close to the average case prediction $\TCVM$ (whether this
is only a coincidence is still not clear). Most importantly the average
case computation (population dynamics) with the gauge fixed identifies
the same SG critical temperature $T_\text{CVM-GF} \simeq 1.28$ measured
on single samples (where $T_\text{SG} \simeq 1.27$).

Finally we compared the fixed point solutions found by the GBP message
passing with those found by the provably convergent Double-Loop
algorithm and the message passing heuristic of the Two-Ways algorithm
of \cite{HAK03}. All the algorithms find the same paramagnetic
solutions at high T, while below $T_\text{SG}$ they find a spin glass
solution, in the sense that local magnetizations are non zero, while
the global magnetization is null.  Decreasing the temperature
Double-Loop and HAK switch back from the spin glass to the
paramagnetic solution, at the cost of a factor $10^2-10^3$ and
$10-10^2$ respectively in running time, compared to GBP. Furthermore,
the paramagnetic solution can always be found fast by the Dual
algorithm of \cite{dual}, making these two algorithms (Double-Loop and
HAK) unnecessarily slow.

Although the thermodynamics of the 2D EA model is paramagnetic, at low
temperatures, the correlation length grows until eventually surpassing
$L/2$ and therefore being effectively infinite for any finite size 2D
system. In such a situation the non paramagnetic solutions obtained by
GBP can account for long range correlations, and presumably gives
better estimates for the correlations among spins than the
paramagnetic solution obtained by HAK and Double Loop.

Establishing the previous claim requires a detailed study of the
quality of CVM approximation at low temperatures (in the non
paramagnetic range) and its connections to the statics and dynamics of
2D Edwards Anderson model, which is already under study. Application
of CVM and GBP message passing to Edwards Anderson model in 3D is also
appealing, since this model does have a spin glass behavior at low
temperature.

\bibliography{bibliografia}

\begin{thebibliography}{32}
\bibitem [{def()}]{defTbethe}%
  \BibitemOpen
  \href@noop {} {}\bibinfo {note} {The Bethe temperature
  $T_{\scriptstyle{Bethe}}$ is the one at which a non trivial spin glass
  solution appears for a random regular Bethe lattice with connectivity $K=4$.
  The Bethe lattice looks locally as a tree.}\BibitemShut {Stop}%
\end{thebibliography}

\end{document}